\documentstyle[aps,prl,twocolumn,graphicx]{revtex}

\newcommand{\be}{\begin{eqnarray}}
\newcommand{\ee}{\end{eqnarray}}

\begin{document}
\draft
\twocolumn[\hsize\textwidth\columnwidth\hsize\csname @twocolumnfalse\endcsname
\title{Thermal/Electronic Transport Properties and Two-Phase Mixtures in La$_{5/8-x}$Pr$_x$Ca$_{3/8}$MnO$_{3}$}
\author{K. H. Kim$^{1,*}$, M. Uehara$^{1,\dagger}$, C. Hess$^{1,\ddagger}$,%
P. A. Sharma$^{1}$ and S-W. Cheong$^{1,2}$}
\address{$^1$Department of Physics and Astronomy, Rutgers University, Piscataway, New
Jersey 08854}
\address{$^2$Bell Laboratories, Lucent Technologies, Murray Hill, New Jersey 07974}
\date{to appear in Phys. Rev. Lett.}
\maketitle

\begin{abstract}
We measured thermal conductivity $\kappa $, thermoelectric power {\it S}, and dc
electric conductivity $\sigma$ of La$_{5/8-x}$Pr$_{x}$Ca$_{3/8}$MnO$_{3}$, 
showing an intricate interplay
between metallic ferromagnetism (FM) and charge ordering (CO) instability.
The change of $\kappa $, {\it S} and $\sigma$ with temperature ({\it T}) and {\it x} agrees well with the
effective medium theories for binary metal-insulator mixtures. This
agreement clearly demonstrates that with the variation of {\it T} as well as {\it x},
the relative volumes of FM and CO phases drastically change and percolative
metal-insulator transition occurs in the mixture of FM and CO domains.
\end{abstract}
 \vskip0.5pc] 
\newpage

Immense resurgent activities on mixed-valent manganites reveal the
importance of Jahn-Teller-type electron/lattice coupling in addition to the
double exchange mechanism\cite{DE,Millis,Zhao,OO}. Another important aspect
of manganites is the existence of various-scale, real-space variation of
physical properties or parameters. For example, even though the
orthorhombicity of the average structure drastically decreases with the
replacement of La by divalent ions in LaMnO$_{3}$, the large variation of
local Mn-O bond lengths still remains intact\cite{Louca}. In high
divalent-ion doping ranges, charge ordering results in sheet-like
arrangements of Mn$^{3+}$ and Mn$^{4+}$ ions with 10-30 \r{A} length-scales 
\cite{Cheong}. Furthermore, ferromagnetic resonance experiments have shown
the presence of two types of signals in ferromagnetic manganites, which was
interpreted as evidence of electronic phase separation \cite{Allodi}.
Various experiments suggest the existence of magnetic polarons or mobile
ferromagnetic clusters at high temperature ({\it T}), which could be viewed
as resulting from dynamic phase separation \cite{Zhou,Teresa}. Recently, an
electron diffraction study on low Curie temperature ({\it T}$_{\text{C}}$)
manganites has revealed that there coexist ferromagnetic (FM)-metallic and
CE-type charge-ordered (CO), insulating domains \cite{Uehara,CE}. It has
been emphasized that this particular type of static phase separation is
responsible for colossal magnetoresistance in low {\it T}$_{\text{C}}$
manganites. Various theoretical models for mixed-valent manganites also
reveal the general tendency of static or dynamic electronic phase separation 
\cite{Yunoki}.

The transport properties of metal-insulator (M-I) mixtures have been
perennial topics for both theoretical and experimental condensed matter
physics \cite{Inhomogeneous}. Most of the experiments were performed on
films with deposited M-I mixture \cite{Hurvits} or artificial bulk M-I
composites prepared under pressure \cite{Ahn}. The total electric and
thermal conductivity and thermoelectric power of binary M-I mixtures were
successfully explained by the effective medium theories \cite
{Inhomogeneous,Hurvits,Ahn,GEM,Dupez,Bergman}.

In this letter, we report the absolute values of the magnetization {\it M},
thermal conductivity $\kappa $, thermoelectric power {\it S}, and dc
electric conductivity ${\it \sigma }$ of La$_{5/8-x}$Pr$_{x}$Ca$_{3/8}$MnO$%
_{3}$ with various {\it x} and {\it T}. Various aspects of our results are
consistent with the coexistence of FM and CO phases, whose relative volumes
change with both {\it T} and {\it x}, and the percolative M-I transition in
FM-CO mixtures. Furthermore, the {\it T} and {\it x} dependence of ${\it %
\sigma }$, ${\it \kappa}$, and {\it S} agrees well with the (general)
effective medium theories for M-I mixtures.

High-quality polycrystalline specimens of La$_{5/8-x}$Pr$_{x}$Ca$_{3/8}$MnO$%
_{3}$ with {\it x}=0.0, 0.1, 0.2, 0.25, 0.3, 0.35, 0.375, 0.40, 0.42, and
0.625 have been prepared with the standard solid state reaction. We fixed
the Ca concentration at 3/8 because our previous studies showed that {\it T}$%
_{\text{C}}$ is optimized at this particular Ca doping level \cite{Cheong}. $%
{\it \sigma }$ of all specimens with accurate geometry was measured with the
standard four probe method, and {\it M} was measured with a SQUID
magnetometer. Both ${\it \kappa }$ and ${\it S}$ of the representative
samples ({\it x}=0.0, 0.1, 0.25, 0.35, 0.375, 0.42, and 0.625) have been
measured from 8 to 310 K with the steady state method. A radiation shield
was used to obtain absolute ${\it \kappa }$ values\cite{grain}.

The systematic {\it T}-dependent and {\it M/H} curves are shown in Fig. 1. 
{\it M/H} curves were measured in {\it H} = 2 kOe, which was carefully
chosen to align FM domains without influencing the CO insulating phase.
These results indicate that La$_{5/8-x}$Pr$_{x}$Ca$_{3/8}$MnO$_{3}$ ({\it x}%
=0.0) is, basically, FM-metallic below 275 K, and that the ground state of Pr%
$_{5/8}$Ca$_{3/8}$MnO$_{3}$ ({\it x}=0.625) is CO-insulating below 225 K.
The behaviors of ${\it \sigma }$ and {\it M/H} for other {\it x} compositions are
systematically in-between those for {\it x}=0.0 and 0.625. Open circles
(Fig. 1 (a)) represent the M-I transition points where ${\it \sigma }$
becomes (local) minimum \cite{x=0.3}. Open circles in Fig. 1 (b) show the 
{\it M/H} values at the M-I transition points determined from the curves.
The average of those {\it M/H} values (dotted line) is 17$\pm $2 \% of 8.1
emu/mol, the saturated {\it M/H} value of {\it x}=0.0. Thus, with changing 
{\it T}, the M-I transition occurs when {\it M} of each sample becomes about
17 \% of that of {\it x}=0.0, independent from {\it x} value. If we assume
that the ${\it T}$-dependent volume fraction {\it f}(${\it T}$) of the FM
domain is proportional to {\it M}(${\it T}$) in H=2 kOe, the M-I transition
with changing ${\it T}$ occurs when {\it f} reaches $\sim $0.17, close to
the three-dimensional percolation threshold ({\it f}$_{\text{c}}$). This
behavior can be also seen in the variation of ${\it \sigma }$ with {\it x}
for fixed ${\it T}$. In Fig. 2 (a), ${\it \sigma }$ vs. {\it M}$_{{\it x}}$($%
{\it T}$)/{\it M}$_{\text{0.0}}$(${\it T}$) plot is shown at 10 and 100 K
for various {\it x}. Interestingly, the samples with {\it M}$_{{\it x}}$($%
{\it T}$)/{\it M}$_{\text{0.0}}$(${\it T}$) $<$ 0.15 are in the insulating
regions of the curves in Fig. 1, while those with {\it M}$_{{\it x}}$(${\it T%
}$)/{\it M}$_{\text{0.0}}$(${\it T}$) $>$ 0.17 are in the metallic regimes.
Therefore, with variation of {\it x}, the M-I transition occurs when {\it M}$%
_{{\it x}}$/{\it M}$_{0.0}\approx $0.15-0.17. These observation clearly
suggest that the M-I transition with both ${\it T}$ and {\it x} takes place
when {\it f}(${\it T}$,{\it x}) ($\equiv ${\it M}$_{{\it x}}$(${\it T}$)/%
{\it M}$_{\text{0.0}}$(${\it T}$)) becomes close to {\it f}$_{\text{c}}$. It
is important to note that {\it M}$_{0.0}$(${\it T}$) contains the natural $%
{\it T}$-dependence of ferromagnetic moment so that {\it f}(${\it T}$,{\it x}%
) equals to {\it M}$_{{\it x}}$(${\it T}$)/{\it M}$_{\text{0.0}}$(${\it T}$)
(not {\it M}$_{{\it x}}$(${\it T}$)/{\it M}$_{\text{0.0}}$(${\it T}$=0)).

To gain further insights into the nature of the M-I transition, we measured $%
{\it T}$-dependent $\kappa $ and ${\it S}$ as shown in Fig. 3. First, we
note that the estimated $\kappa _{e}$ from ${\it \sigma }$, using the
Wiedemann-Franz law, is by two orders of magnitude smaller than the measured 
$\kappa $(${\it T}$) for all {\it x}, indicating the dominant phonon
contribution. Furthermore, at high ${\it T}$ above ${\it T}_{_{\text{C}}}$
or ${\it T}_{_{\text{CO}}}$, $\kappa $ always increases when ${\it T}$ is
raised, and the magnitude of $\kappa $ is in the range of 0.5-2 W/mK,
comparable to that of amorphous solids \cite{Berman}. This behavior has been
attributed to local anharmonic lattice distortions associated with small
polarons \cite{TC}. Related to this, ${\it S}$(${\it T}$) of all the samples
at high ${\it T}$ follows the form {\it S}$_{0}$+{\it E}$_{g}$/{\it k}$_{B}%
{\it T}$ with the gap energy {\it E}$_{g}$ systematically increasing from 4
meV ({\it x}=0.0) to 12 meV ({\it x}=0.625). These {\it E}$_{g}$ values are
significantly smaller than the activation energies (125 meV: {\it x}=0.0 to
175 meV: {\it x}=0.625) associated with ${\it \sigma }$(${\it T}$). This difference can
result from the small polaronic transport \cite{TP}.

As shown in Fig. 3 (a), with increasing {\it x}, $\kappa $(${\it T}$)
smoothly evolves from that of {\it x}=0.0 to that of {\it x}=0.625, and the $%
\kappa $ increase at ${\it T}_{\text{C}}$ becomes smaller. Consistently, $%
\kappa $ vs. {\it M}$_{{\it x}}$(${\it T}$)/{\it M}$_{\text{0.0}}$(${\it T}$%
) at 10 and 100 K (Fig. 2 (b)) shows that $\kappa $ varies monotonically
from the maximum ($\kappa $ of {\it x}=0.0) to the minimum ($\kappa $ of 
{\it x}=0.625). For {\it x}=0.0, $\kappa $ at ${\it T}_{\text{C}}$ increases
suddenly, probably due to the suppression of local lattice distortions
associated with small polarons. For {\it x}=0.625, $\kappa $(${\it T}$)
behaves as in amorphous solids in the entire ${\it T}$ range, except for a
slight increase at ${\it T}_{\text{co}}$.

In comparison with $\kappa $ and ${\it \sigma }$, ${\it S}$ exhibits seemingly different
behaviors with {\it x}. ${\it S}$ is very close to the metallic value, ${\it %
S}$ of {\it x}=0.0, even near {\it f}$_{\text{c}}$ where ${\it \sigma }$ (or 
$\kappa $) is still significantly smaller than ${\it \sigma }$ (or $\kappa $%
) of {\it x}=0.0. ${\it S}$ vs. {\it M}$_{{\it x}}$/{\it M}$_{{\it 0.0}}$ at
100 K (Fig. 2 (c)) clearly demonstrates this tendency; ${\it S}$ is close to
zero (slightly negative), and is insensitive to {\it M}$_{{\it x}}$/{\it M}$%
_{\text{0.0}}$ as long as {\it M}$_{{\it x}}$/{\it M}$_{\text{0.0}}$ $\sim $%
10 \%. In contrast, ${\it \sigma }$ for {\it M}$_{{\it x}}$/{\it M}$_{\text{%
0.0}}$ = 10 \% at 100 K is more than three orders of magnitude smaller than
that of {\it x}=0.0. In fact, ${\it T}$-dependence of ${\it S}$ near ${\it T}%
_{\text{C}}$ is also consistent with this metallic ${\it S}$ behavior near 
{\it f}$_{\text{c}}$. With decreasing ${\it T}$ near ${\it T}_{\text{C}}$, $%
{\it S}$ starts to decrease, i. e., becomes metallic at ${\it T}$ higher
than those for $\kappa $ and ${\it \sigma }$ changes. For example, in the
heating curves of {\it x}=0.35, ${\it S}$ starts to decrease around 130 K,
significantly higher than ${\it T}$ (100 K) for abrupt $\kappa $ increase or 
${\it T}$ ($\sim $110 K) for ${\it \sigma }$ minimum.

When thermal/electronic transport properties of our systems are viewed as
those of M-I mixtures, the above peculiar ${\it S}$ behavior is, in fact,
consistent with the theoretical prediction of effective thermoelectric power 
${\it S}_{\text{E}}$ by Bergmann and Levy \cite{Bergman}. For an isotropic
binary mixture, they showed that in terms of ${\it \sigma }$, $\kappa $, and 
${\it S}$ of each component,$\ {\it S}_{\text{E}}$ is given by 
\begin{equation}
S_{\text{E}}=S_{\text{M}}+(S_{\text{I}}-S_{\text{M}})\left( \frac{\kappa _{%
\text{E}}/\kappa _{\text{M}}}{\sigma _{\text{E}}/\sigma _{\text{M}}}%
-1\right) /\left( \frac{\kappa _{\text{I}}/\kappa _{\text{M}}}{\sigma _{%
\text{I}}/\sigma _{\text{M}}}-1\right) \text{ },\ 
\end{equation}

where the subscripts M and I refer to metallic and insulating components,
respectively. ${\it \kappa }_{\text{E}}$ and ${\it \sigma }_{\text{E}}$
refer to effective thermal and electric conductivity, respectively, of the
binary mixture. This equation has been successfully applied to explain ${\it %
S}$ behaviors of binary Al-Ge films \cite{Hurvits}. When ${\it \sigma }_{%
\text{I}}$/${\it \sigma }_{\text{M}}$ $<$$<$ $\kappa _{\text{I}}$ /$\kappa _{%
\text{M}}<$1 (which applies to our system) and for {\it f} = {\it f}$_{\text{%
c}}$ , the above equation leads to$\ {\it S}_{\text{E}}\approx {\it S}_{%
\text{M}}$, which explains our experimental results noted above.

To quantitatively compare our results with Eq. (1), we calculated$\ {\it S}_{%
\text{E}}$ at every ${\it T}$ \cite{Previous}. In this comparison,
experimental ${\it \sigma }$ and $\kappa $, shown in Figs. 1 and 3, were
used for ${\it \sigma }_{\text{E}}$ and ${\it \kappa }_{\text{E}}$ , and ($%
\sigma _{\text{I}}$, $\kappa _{\text{I}}$, and ${\it S}_{\text{I}}$) and ($%
\sigma _{\text{M}}$, $\kappa _{\text{M}}$, and$\ {\it S}_{\text{M}}$) are
assumed to be identical with those of {\it x}=0.0 and 0.625, respectively.
(For large {\it x}, it was difficult to measure ${\it S}$ at low ${\it T}$
due to high resistivity, so we assumed that ${\it S}_{\text{I}}$ changes as
1/${\it T}$, and that ${\it \sigma }_{\text{I}}$ does exponentially.) The
solid lines in Fig. 3 (b) depict the calculated$\ {\it S}_{\text{E}}$, using
Eq. (1), for heating curves of {\it x}=0.25, 0.35, and 0.42. The calculated
curves match with our experimental ${\it S}$ surprisingly well at all ${\it T%
}$ below ${\it T}_{\text{C}}$ or ${\it T}_{\text{co}}$.

For ${\it \sigma }_{\text{E}}$ (or ${\it \kappa }_{\text{E}}$) of a binary
M-I mixture, Mclachlan \cite{GEM} proposed the general effective medium
(GEM) equation, 
\begin{equation}
(1-f)\left( \frac{\sigma _{I}^{1/t}-\sigma _{E}^{1/t}}{\sigma
_{I}^{1/t}+A\sigma _{E}^{1/t}}\right) +f\left( \frac{\sigma
_{M}^{1/t}-\sigma _{E}^{1/t}}{\sigma _{M}^{1/t}+A\sigma _{E}^{1/t}}\right) =0%
\text{ },
\end{equation}
where $A$ =(1-$f_{\text{c}}$)/$f_{\text{c}}$. The same equation also works
for $\kappa $. The critical exponent $\ t$ is close to 2 in three dimension.
This equation has been successfully applied to isotropic inhomogeneous media
in wide {\it f} regions including percolation regime \cite{Hurvits,GEM,Dupez}%
.

To apply the GEM equation to ${\it \sigma }$(${\it T}$), we assumed that 
{\it f}(${\it T}$,{\it x}) = {\it M}$_{{\it x}}$(${\it T}$)/{\it M}$_{\text{%
0.0}}$(${\it T}$), ${\it \sigma }_{\text{M}}$(${\it T}$) = ${\it \sigma }$($%
{\it T}$) of {\it x}=0.0, and ${\it \sigma }_{\text{I}}$(${\it T}$) = ${\it %
\sigma }$(${\it T}$) of {\it x}=0.625. With the parameters $\ t$=2 \& {\it f}%
$_{\text{c}}$=0.17, the calculated$\ {\it \sigma }_{\text{E}}$ for various 
{\it x} are shown as solid lines in Fig. 4. At ${\it T}$ $>$ $\sim $80 K,$\ 
{\it \sigma }_{\text{E}}$(${\it T}$) nicely matches the experimental ${\it %
\sigma }$(${\it T}$) even if ${\it \sigma }$ changes by 6 orders of
magnitude with ${\it T}$ and {\it x}. However, this agreement does not hold
at very low ${\it T}$. The calculated$\ {\it \sigma }_{\text{E}}$(${\it T}$)
at ${\it T}$ $<$ 80 K with the same parameters $\ t$=2 \& {\it f}$_{\text{c}%
} $=0.17 significantly deviated from the experimental ${\it \sigma }$(${\it T%
}$). We found that at ${\it T}$ $<$ 80 K, the calculated$\ {\it \sigma }_{%
\text{E}}$(${\it T}$) matches the experimental ${\it \sigma }$(${\it T}$)
better when $\ t$ is increased to $\sim $ 4. This is more evident in the 
{\it x} dependence of ${\it \sigma }$ at 10 and 100 K, as shown in Fig. 2
(a).$\ {\it \sigma }_{\text{E}}$(100 K), calculated with $\ t$=2 \& {\it f}$%
_{\text{c}}$=0.17 (solid line), matches the experimental values (open
circles) better than that with $\ t$=4 \& {\it f}$_{\text{c}}$=0.17 (dotted
line). However,$\ {\it \sigma }_{\text{E}}$(10 K), calculated with {\it $\ t$%
}=4 \& {\it f}$_{\text{c}}$=0.15 (solid line), is closer to the experimental 
${\it \sigma }$ (solid circles) than that with $\ t$=2 \& {\it f}$_{\text{c}%
} $=0.15 (dashed line) \cite{LowT}. (The change of {\it f}$_{\text{c}}$ in
the range of 0.15-0.17 makes little difference.) These observations
demonstrate that $\ t$, normally close to the three dimensional exponent of
2, becomes $\sim $4 at very low ${\it T}$. A similar, drastic increase of $\
t$ has been noted in the case of tunneling transport for M-I mixtures \cite
{Hurvits,Balberg}, suggesting that the tunneling process between FM domains
is important for ${\it \sigma }$ of our system at very low ${\it T}$ \cite
{Hwang}.

By using the GEM equation for ${\it \kappa }_{\text{E}}$, the $\kappa $($%
{\it T}$) for various {\it x} can be calculated with the assumption that $\
\kappa _{\text{I}}$(${\it T}$)=$\kappa $(${\it T}$) of {\it x}=0.625, $\
\kappa _{\text{M}}$(${\it T}$)=$\kappa $(${\it T}$) of {\it x}=0.0, $\ \ t$%
=2, and {\it f}$_{\text{c}}$=0.17. The solid lines in Fig. 3 (a) represent
the estimated ${\it \kappa }_{\text{E}}$ for {\it x}=0.1, 0.25, 0.35, and
0.42. In addition, the calculated ${\it \kappa }_{\text{E}}$ as a function
of {\it M}$_{{\it x}}$/{\it M}$_{\text{0.0}}$ at 10 and 100 K is depicted as
solid lines in Fig. 2 (b). Estimated ${\it \kappa }_{\text{E}}$ lines in
Figs. 2 and 3 coincide with the experimental data well \cite{t}. In order to
confirm self-consistency,$\ {\it S}_{\text{E}}$ at 10 and 100 K is evaluated
by using the calculated$\ {\it \sigma }_{\text{E}}$ and ${\it \kappa }_{%
\text{E}}$ (solid lines of Fig. 2 (a) and (b)), and the Eq. (1). The
calculated$\ {\it S}_{\text{E}}$ (solid lines of Fig. 2 (c)) with the
variation of {\it x} is in good agreement with the experimental values.

The unambiguous agreement between the measured thermal/electronic transport
properties and the calculated values based on Eqs. (1) and (2) strongly
indicates that: (1) transport properties are dominated by thermal/electrical
conduction in M-I mixtures, (2) the relative volume of the (FM) metallic
phase is proportional to the measured {\it M}(${\it T}$,{\it x}), and (3)
the ${\it T}$-dependent transport and magnetic properties of metallic and
insulating phases are always that of {\it x}=0 and 5/8, respectively.
Combined with the earlier electron diffraction results\cite{Uehara}, this
successful agreement demonstrates that all of the thermal, electronic, and
magneto-transport properties of La$_{5/8-{\it x}}$Pr$_{{\it x}}$Ca$_{3/8}$MnO%
$_{3}$ are dominated by the percolative conduction through FM-metallic
domains which is statically mixed with CO insulating domains. One surprising
indication from our results is that at least in low ${\it T}_{C}$ materials (%
{\it x} $>$ 0.25), the so-called Curie transition is, in fact, the M-I
transition across percolation threshold, and the ordered FM moment changes
smoothly near the percolative phase transition ${\it T}$.

We greatly thank A. J. Millis, G. Kotliar, E. Abrahams, and T. W. Noh for
useful discussions. We are partially supported by the NSF-DMR-9802513. K. H.
Kim and M. U. are partially supported by the KOSEF and by the JPSJ
Fellowship, respectively.

\newpage
\begin{figure}[tbp]
\caption{(a) $\protect\sigma $({\it T}) of La$_{5/8-x}$Pr$_x$Ca$%
_{3/8}$MnO$_3 $ for cooling (dotted lines) and heating (solid lines). (b) $%
M/H$ curves with zero field cooling. Open circles depict the M-I transition
points determined from the data. {\it T}-dependent volume fraction of the FM
domains, {\it f}({\it T}), for each {\it x}, was determined as {\it f}({\it T%
},{\it x}) = $M_x$({\it T})/$M_{0.0}$({\it T}). }
\label{Ref}
\end{figure}

\begin{figure}[tbp]
\caption{{\it S}, $\protect\kappa $, and $\protect\sigma$ values vs. 
$M_{x}$/$M_{0.0} $ at 10 and 100 K. The lines depict the theoretical
predictions by Eqs. (1) and (2). The solid lines in s show the theoretical
results with $\ t $=4 \& $f_c$=0.15 at 10 K and {\it t}=2 \& $f_c$=0.17 at
100 K. The dotted and dashed lines represent the theoretical results with 
{\it t}=2 \& $f_c$=0.15 at 10 K and with $\ t$=4 \& $f_c$=0.17 at 100 K,
respectively. For the theoretical predictions (solid lines) of $\protect%
\kappa $ at 10 and 100 K, {\it t}=2 \& $f_c$=0.17 were used.}
\label{Sig}
\end{figure}

\begin{figure}[tbp]
\caption{(a) $\protect\kappa $({\it T}) for cooling (crosses) and heating
(solid circles). The solid lines show the predictions of $\protect\kappa $($%
{\it T}$) (heating) by Eq. (2) with {\it t}=2 \& $f_c$=0.17. (b) {\it S}(%
{\it T}) of La$_{5/8-x}$Pr$_{x}$Ca$_{3/8}$MnO$_{3}$. Crosses represent
cooling curves for $x$=0.42 and 0.35, and the other symbols represent
heating data. ({\it S} of {\it x}=0.1 is very close to that of {\it x}=0.0
and 0.25 below $T_C$ and omitted for clarity.) The solid lines show the
predictions of Eq. (1) for {\it x}=0.25, 0.35, and 0.42 with heating. }
\label{TOPeak}
\end{figure}

\begin{figure}[tbp]
\caption{Theoretical predictions (solid lines) of $\protect\sigma$(%
{\it T}) by the GEM equation (Eq. (2)) with {\it t}=2 \& $f_c$=0.17 above $%
\sim$80 K. Solid squares are the experimental data redrawn from Fig. 1. }
\label{XRD}
\end{figure}

\newpage
\clearpage
\vspace*{0cm}
\begin{center}
\hspace*{0cm}
\includegraphics[height=22cm,width=17cm,angle=0]{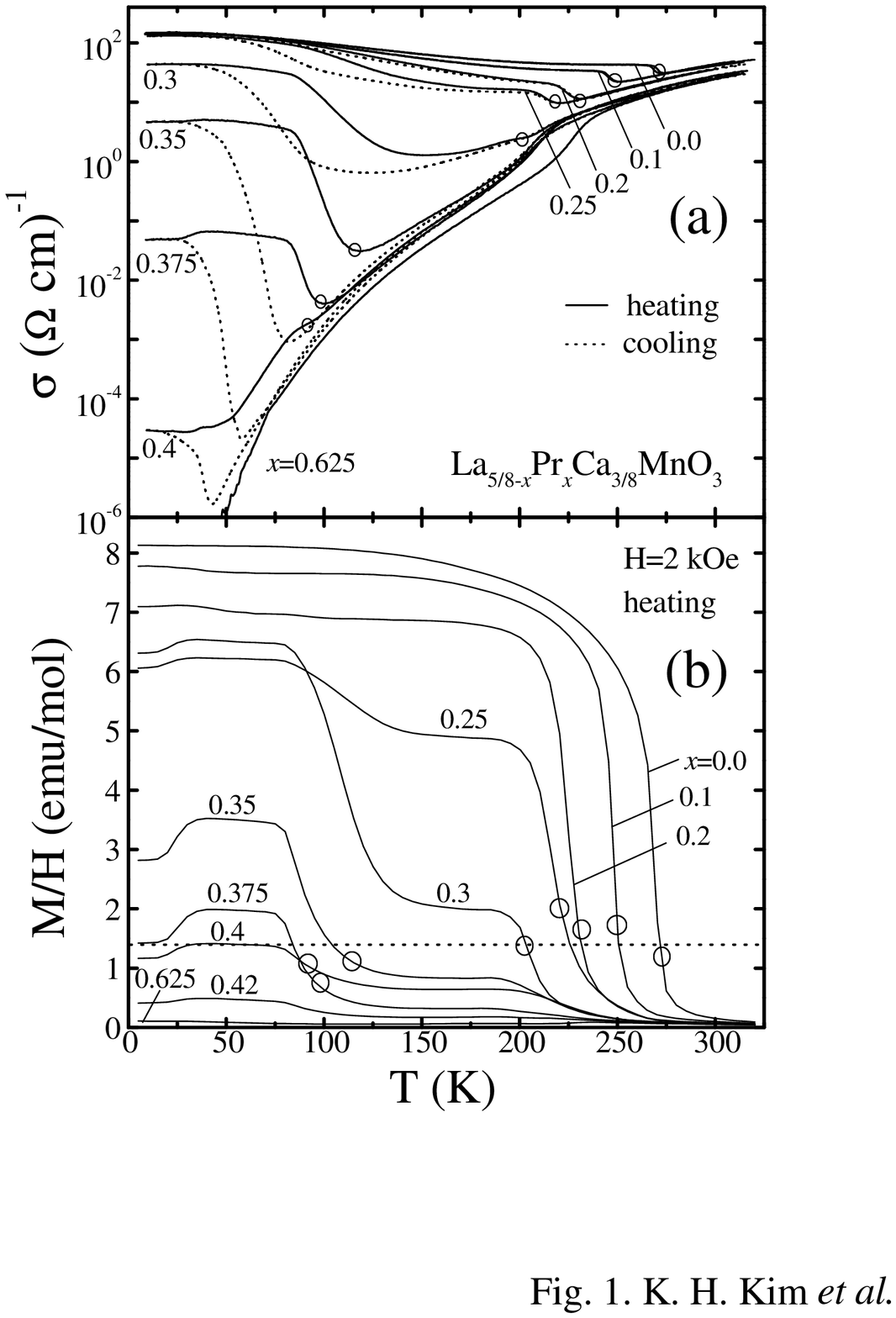}
\end{center}

\newpage
 \clearpage
\vspace*{0cm}
\begin{center}
\hspace*{0cm}
\includegraphics[height=22cm,width=17cm,angle=0]{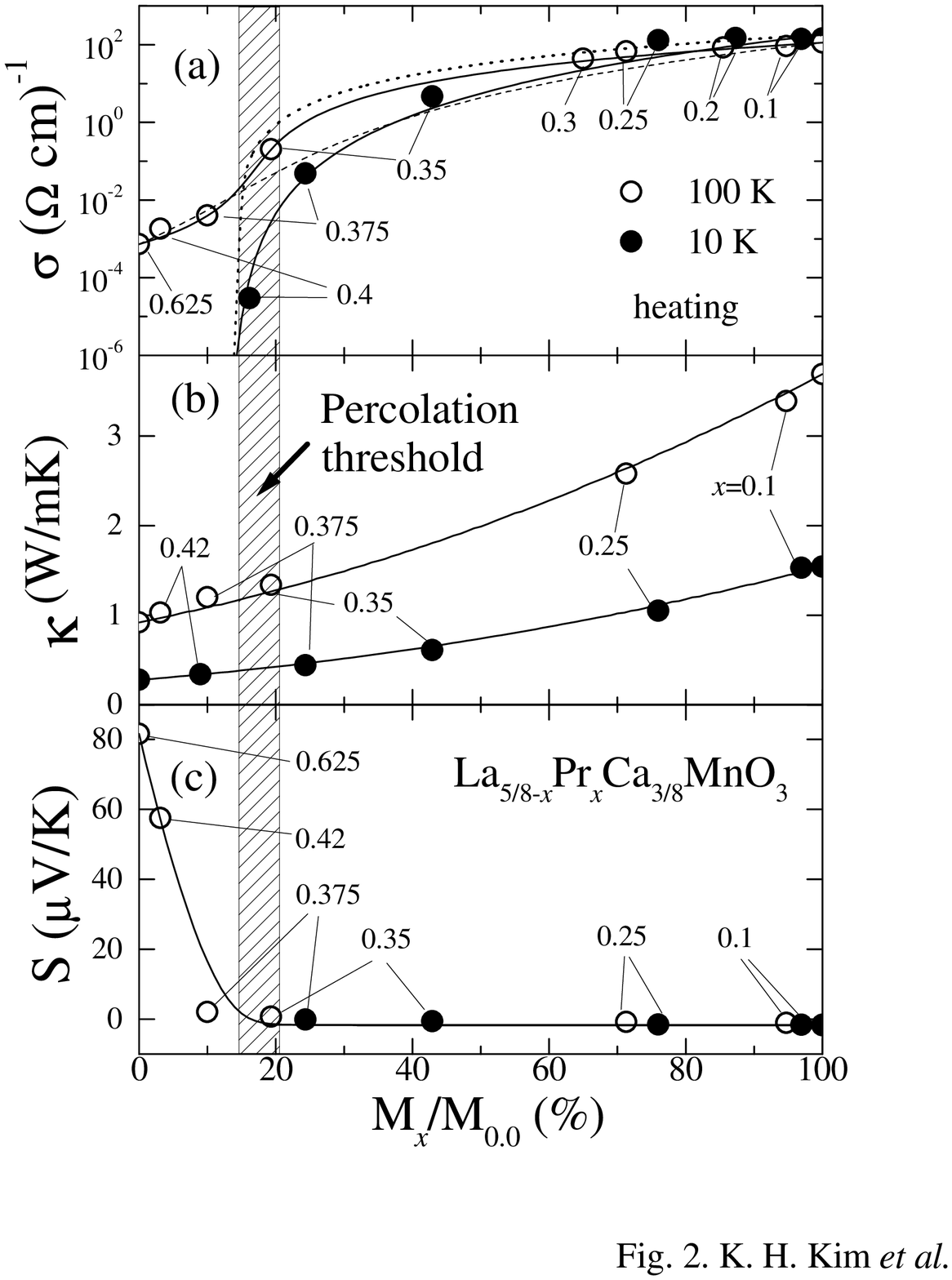}
\end{center}

\newpage
 \clearpage
\vspace*{0cm}
\begin{center}
\hspace*{0cm}
\includegraphics[height=22cm,width=17cm,angle=0]{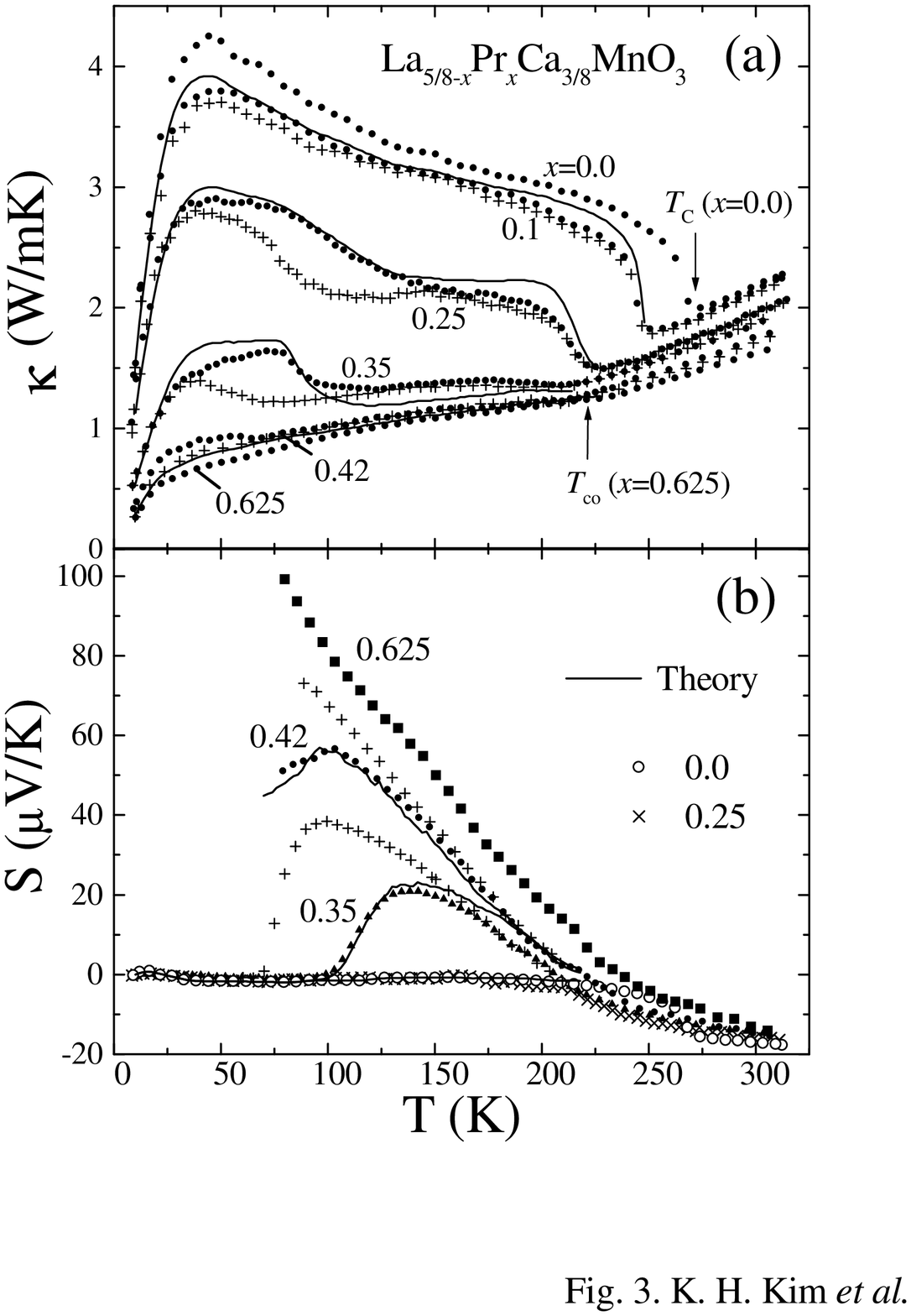}
\end{center}
\vspace*{0.0cm}

\newpage
 \clearpage
\vspace*{0cm}
\begin{center}
\hspace*{0cm}
\includegraphics[height=17cm,width=21cm,angle=-90]{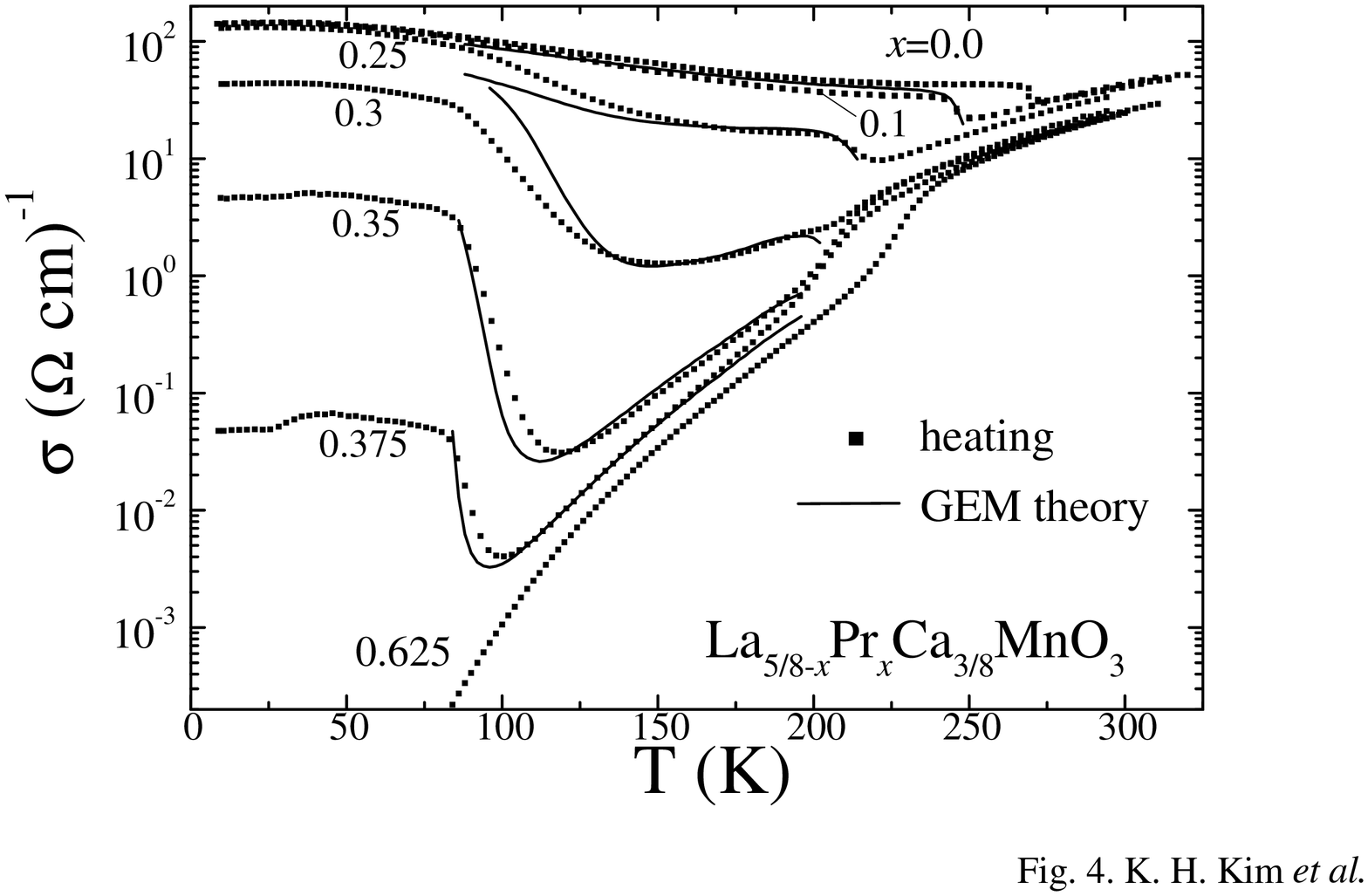}
\end{center}
\vspace*{0.0cm}


\begin{references}
\bibitem{grain}  To reduce the complexity due to grain boundaries on the
measured $\kappa $ and $\sigma $, we used the same, optimized sythesis
conditions for all the samples.
\end{references}
\end{document}